\documentclass[conference]{IEEEtran}
\usepackage{amsmath}
\usepackage{amssymb}
\usepackage{graphicx}
\usepackage{cite}
\usepackage{algorithm}
\usepackage{algorithmicx}

\usepackage{algpseudocode}
\usepackage[square, comma, sort&compress, numbers]{natbib}
\usepackage{balance}
\usepackage{color}

\usepackage{algpseudocode}
\usepackage{url}

\usepackage{amsfonts}
\usepackage{array}
\usepackage{amssymb}
\usepackage{amsmath}
\usepackage{amssymb}
\usepackage{tabularx}
\usepackage{mathrsfs}
\usepackage{amsthm}
\usepackage{multirow}
\usepackage{bpchem}
\usepackage{subfigure}
\usepackage{float}
\usepackage[justification=centering]{caption}

\usepackage{array,multirow,makecell}

\ifCLASSINFOpdf
\else
\fi

\usepackage{array}

\begin{document}
%
\title{Urban Rail Transit System Operation Optimization: A Game Theoretical Methodology}

\author{\IEEEauthorblockN{Jiao Ma$^1$, Changle Li$^{1,2,*}$, Weiwei Dong$^1$, Zhe Liu$^1$, Tom H. Luan$^3$, Lina Zhu$^1$, and Lei Xiong$^2$}
\IEEEauthorblockA{$^1$State Key Laboratory of Integrated Services Networks, Xidian University, Xi'an, Shaanxi, 710071 China}
\IEEEauthorblockA{$^2$State Key Laboratory of Rail Traffic Control and Safety, Beijing Jiaotong University, Beijing, 100044 China}
\IEEEauthorblockA{$^3$School of Information Technology, Deakin University, Melbourne, VIC, 3125 Australia}
$^*${clli@mail.xidian.edu.cn}}


\maketitle

\begin{abstract}
The Urban Rail Transit (URT) has been one of the major trip modes in cities worldwide. As the passengers arrive at variable rates in different time slots, e.g., rush and non-rush hours, the departure frequency at a site directly relates to perceived service quality of passengers; the high departure frequency, however, incurs more operation cost to URT. Therefore, a tradeoff between the interest of railway operator and the service quality of passengers needs to be addressed. In this paper, we develop a model on the operation method of train operation scheduling using a Stackelberg game model. The railway operator is modeled as the game leader and the passengers as the game follower, and an optimal departure frequency can be determine the tradeoff between passengers' service quality and operation cost. We present several numerical examples based on the operation data from Nanjing transit subway at China. The results demonstrate that the proposed model can significantly improve the traffic efficiency.
%
\end{abstract}


\IEEEpeerreviewmaketitle

\section{Introduction}
The world has witnessed great development and rapid improvement of urban rail transit (URT) due to its convenience, safety, comfort, vast capacity and high energy efficiency. In China, for example, 88 urban rail lines had been built with over 3000 kilometers of track operation by the end of 2014. Another 1400 kilometers had been added with a total investment of \$170 billion USD by 2015~\cite{1}. It is foreseeable that in the forthcoming future, the modern urban rail transit system will be the backbone transportation system in the large and medium-size cities.

With increasing people relying on the subway as their commutes in cities, challenges are imposed on the service operators of URT to maintain an acceptable or satisfying level of quality of experience (QoE)~\cite{2, 3}. Taking the departure frequency as an example, the passengers prefer frequent train arrivals to reduce the waiting time and more spare seats on trains. Passengers, especially from outskirts suburbs, prefer direct line routes with minimal intermediate stops to minimize their trip time. In cases where the demand for transit service is elastic, shorter routes, and thus higher access impedance, might decrease service attractiveness and drive commuters to adopt other travel methods. On the other hand, the cost or interest restriction of operators cannot be ignored. To limit the cost of operation, the operators prefer to a low departure frequency and shorter line routes. Therefore, service operator and passengers have conflicting objectives~\cite{4}.

Apparently, the effective approach to the conflict is to search for the balance between the passengers and operators' interests or demands and design an adaptive and optimized operation mechanism. Supported by the advanced wireless communication technology~[5-7], multi-level information awareness becomes available. By way of example, with the use of wireless sensor networks~[8-10], the passengers counts, the travel time and the load in the vehicle are accessible.

Based on the advanced technology, we introduce a Stackelberg game model~[11-13] to address the conflict of interests. In the game, two parties influence each other involved with their own strategies and interests; therefore, one party adjusts its own combat strategy making use of the other's strategy to achieve the goal. The input of the game model is the benefit expressions of both parties. The output of the game mode is the optimal departure frequency with the interests of both parties addressed. The main contributions of our work are summarized as follows:
\begin{itemize}
  \item \emph{Mathematical modeling}: we derive the benefit expressions of the service operator and passengers. As for the former, we take the difference between income and spending as its benefit, which is involved with the subway fares, energy-consuming cost, maintenance cost, workers' wages and depreciation cost, etc. The waiting time cost and traveling comfort are considered to evaluate the passengers' benefit;
  \item \emph{Game theoretic solution}: different from the existing approaches, we introduce the game theory to model the dynamic decision-making process of departure frequency to obtain the balanced value;
  \item \emph{Simulations}: we conduct extensive simulations to verify the efficiency of our approach. By comparing our approach with previous literature, we show that our proposal can achieve much better performance.
\end{itemize}

The rest of this paper is organized as follows. Section~\ref{section: related works} describes an overview of the decision making and train operation adjustment of urban rail transit. The system model and problem formation is presented in Section~\ref{section: system model and problem formation}. In Section~\ref{section: stackelberg equilibrium}, the Stackelberg model is solved and proved. Then, numerical experiments are conducted in Section~\ref{section: numerical experiments}. Finally, the paper concludes with Section~\ref{section: conclusion}.

\section{Related Works}~\label{section: related works}
In order to maintain a satisfying level of quality of experience, optimization models have been widely adopted to determine the optimal Urban rail transit (URT) train operation with different objectives, such as travel time~\cite{14}, waiting time~\cite{15}, operation cost~[16-18] and robustness~\cite{19}.

Amit \emph{et al.}~\cite{20} apply optimization techniques to solve train timetable optimization problems. Ghoseiri \emph{et al.}~\cite{21} present a multi-objective optimization model for the passenger train scheduling problem on a railroad network. To adjust the arrival/departure times of trains based on a dynamic behavior of demand, Canca \emph{et al.}~\cite{22} develop a nonlinear integer programming model, which can be used to evaluate the train service quality. Considering the user satisfaction parameters, average travel time and energy consumption, Sun \emph{et al.}~\cite{23} present a multi-objective optimization model of the train routing problem. Wang \emph{et al.}~\cite{24, 25} propose a real-time train scheduling model with stop-skipping and solve the problem with the mixed integer nonlinear programming (MINLP) approach and the mixed integer linear programming (MILP) approach. Sun \emph{et al.}~\cite{26} propose an optimization method of train scheduling for metro lines with a train dwell time mode, and lagrangian duality theory is adopted to solve this optimization problem with high dimensionality.

To our knowledge, for most of the optimization models of train scheduling, their planning objectives are constructed from the perspective of the passengers and operator such as train delays and operation cost. However, the previous literature does not consider the complicated interactions between passengers and operator. This paper makes a contribution towards solving this problem by introducing game theoretic model to this optimization problem.

\section{System Model and Problem Formation}~\label{section: system model and problem formation}
This section first justifies assumptions used, and then analyzes the interplays between the railway operator and passengers. Lastly, a Stackelberg game formulation is developed for the optimization problem of operation mechanism.

\subsection{Assumptions}
Urban rail transit system is complicated. To develop an optimization model of the operation mechanism, we first present and justify some assumptions used in the work:

\begin{itemize}
  \item Failures and accidents will not occur. This is a valid assumption as accidents are now very rare in modern URT system.
  \item The flat fare is applied. This means passengers are charged the same fee regardless of the length of the trip, delay or type of service. The assumption can be easily extended to consider trips of different classes and fare rates.
  \item Passengers' arrival rate changes with the departure frequency. This is a working assumption as an increasing departure frequency typically attracts more passengers, and the arrival rate of passengers would increase accordingly;
  \item With new Advanced Public Transportation System technologies (e.g., automatic passenger counters), passengers' arrival rate over stations and time are given or predictable. The estimation of passenger's arrival rate has been an active research problem with many research works developed. Due to the limited space, this paper will not consider the estimation of passenger's arrival rate and assume it known.
\end{itemize}

\subsection{System Model}
As presented above, there is a tradeoff between the operator's and passengers' interests. The passengers would prefer frequent train departures of spare seats and short waiting time. However, the operator targets to high revenue and low operation cost and would control the departure frequency. Consequently, in order to alleviate the perceived conflict between passengers and operator objectives, it is essential and feasible to seek the balance between the two sides' interests. In this paper, a Stackelberg game model is adopted to capture the tradeoff.

In our work, a Stackelberg game with two kinds of players, leaders and followers, is applied into the optimization problem of URT operation mechanism. The leaders have the privilege of acting first and then the followers act according to the leaders' actions sequentially; the players compete on resources to profit in this game by maximizing their own utilities.

In the developed game, the railway operator is the leader and the passengers will follow with correspondent strategies. In specific, as for the adjusting behavior of departure frequency, the operator first imposes a departure frequency $s_1$ from a set $S_1$ according to the traditional method and its cost constraints. Then, informed of the operator's choice, the passengers choose a best budget expenditure on fare $s_2$ from a set $S_2$. In the meantime, the passengers plan their travel routes (which leads to the fluctuation of passenger traffic) according to the predictable travel experience under the current frequency. After observing the response of passengers, the operator readjusts the departure frequency, the former of which can be regarded as a disguised feedback to operator's service. After a continuous adjusting process, the frequency is determined finally. In this process, supervision by public opinion plays an important role who compels the operator to adjust the departure frequency and fare to an appropriate value.

\subsection{Problem Formulation}
\begin{figure}[!h]
  \centering
  \includegraphics[width=3.5in]{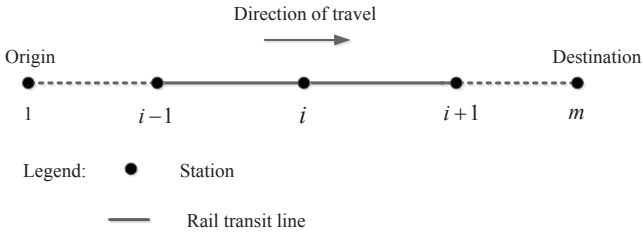}
  \caption{Illustration of a rail transit line. For one transit line, we set up a procedure to obtain the optimization of departure frequency of trains based on the total entrance of passengers into of each station.}\label{line}
\end{figure}

In this paper, we consider train operates on a unidirectional urban transit line with $m$ stations as shown in Fig.~\ref{line}, for bidirectional line can be regarded as two unidirectional lines in our scheme. The stations are numbered as 1, 2, ..., $m$, where station 1 is the origin station and station $m$ is the destination of each trip. Without loss of generality, we assume that the passengers' arrival at these stations follow Poisson distribution. Trains depart from station $1$ to station $m$ in the direction of travel.

\begin{table}[!t]
\renewcommand{\arraystretch}{1.3}
\caption{Important Notations}
\label{table_example}
\centering
\begin{tabular}{cl}
\hline
Notation & Meaning\\
\hline

$i$ & station index, $i = 1,2,...,m$\\

${\lambda _i}$ & passenger arrival rate at station $i$\\

$f$ & departure frequency (per hour)\\

$h$ & operating headway for a transit route\\

$\Omega $ & one-way energy consumption of each train\\

${S_0}$ & daily maintenance cost of each train\\

$R$ & staffs' hourly wage cost \\

$T$ & duration of studied period\\

$\phi $ & depreciation cost of each train\\

${T_p}$ & one-way travel time of each train \\

${T_S}$ & daily operating time of train \\

$\theta$ & rated load factor of train \\

$p$ & passenger capacity of each carriage \\

\hline
\end{tabular}
\end{table}

Based on the scene description above, the utility functions of railway operator and passengers will be analyzed hereinafter.

\subsubsection{Utility Function of Railway Operator (Leader)}
Undoubtedly, the railway operator would like to pursue the greater benefit under the condition that the passengers' basic demands are satisfied. Hence,   it is necessary to establish the utility function of operator.

In this paper, we take the difference between income and operation cost as the benefit. For simplicity, we only take the fare revenue as the operator's income regardless of the advertising revenue, rental income and so on. The income of operator can be expressed as fare $c$ multiplied by the total passenger traffic ${Q^{'}}$ as~(\ref{equ:ECS1}). The actual total passenger traffic ${Q^{'}}$ is influenced by fare $c$ and departure frequency $f$. The impact factors of $c$ and $f$ are ${e_c}$ and ${e_f}$, respectively. The increase of fare $c$ leads to the reduction of passenger traffic while the increase of frequency may attract more passengers.

\begin{equation}
\begin{aligned}\label{equ:ECS1}
    I{\rm{ = c}}{Q^{'}}{\rm{ = c}}Q\left( {1 - {e_c}c + {e_f}f} \right).
\end{aligned}
\end{equation}

The operation cost consist of fixed cost (such as construction cost, labor cost,\emph{ etc}) and variable cost. Considering that the fixed cost do not change along the departure times, we neglect this part of cost. The variable cost we considered include energy-consuming cost, maintenance cost and depreciation cost, etc. Energy consuming cost can be calculated as $\Omega Tf$. Here $\Omega $ is the energy consumption of each train for one trip, $T$ is duration of the studied period, and $f$ is departure frequency. Maintenance cost and depreciation cost can be calculated by
$2{T_p}f  \left( {\phi  + RT + {S_0}\frac{T}{{{T_S}}}} \right)$. The total operation cost is
\begin{equation}
\begin{aligned}\label{equ:ECS2}
    O=\Omega Tf + 2{T_p}f  \left( {\phi  + RT + {S_0}\frac{T}{{{T_S}}}} \right).
\end{aligned}
\end{equation}

We define the utility function of railway operator as

\begin{equation}
\begin{aligned}\label{equ:ECS3}
    {U_l} = & I - O.
\end{aligned}
\end{equation}



To maximize the net income of railway operator, we formulate the optimization problem of operator as follows:

\begin{equation}
\begin{aligned}\label{equ:ECS4}
    &\mathop {\max }\limits_{f,c} {\rm{ }}\indent{U_l}\\
    &{\rm{s}}{\rm{.t}}{\rm{.      }}\indent{\theta _0}{\rm{  < }}\theta  < 1\\
    &{\rm{         }}\indent{f_{\min }}{\rm{  < }} f < {\rm{ }}{f_{\max }}.
\end{aligned}
\end{equation}

The operation of rail transit system needs to satisfy the demand of passengers as well as the operation efficiency. For a given rail transit system, $\theta$ donates the load factor of train, which reflects the utilization of the maximum capacity. $\theta$ should satisfy ${\theta _0}{\rm{  < }}\theta  < 1$ and can be formulated as

\begin{equation}
\begin{aligned}\label{equ:ECS5}
    \theta {\rm{ = }}\frac{Q^{'}}{C}{\rm{ = }}\frac{Q^{'}}{{f \times l \times p}},
\end{aligned}
\end{equation}

in which $C$ means the capacity of the rail transit system. $l$ is the number of carriages on one train, and $p$ is the maximum carriage capacity of each carriage. ${\theta _0}$ is the minimum load factor that operator can accept. $\theta$ is set to be 0-100\%. When $\theta$ equals to 1, it means that the congestion of the carriage reaches the maximum, and this is intolerable. When $\theta$ is close to 0, it means that the carriage is very empty, and the transport resources are wasted. To achieve higher operation efficiency, $\theta$ must be within an appropriate range. ${f_{\min }}$ and ${f_{\max }}$ are the minimal and maximal departure frequency of train, respectively. To ensure driving safety, the maximal frequency can be calculated based on the speed limited, grade profiles and train dynamics. The minimal frequency depends on passengers' demand.
\subsubsection{Utility Function of Passengers (Follower)}
Passengers benefit from accessing the traveling service provided by railway operator, and pay the operator fare. Given the departure frequency $f$, passengers would determine the fare they are willing to pay for their received service, by considering both the gained QoE and incurred expense. The utility of passengers has the following characteristic:
\begin{itemize}
  \item The utility of passengers is a concave function of the fare $c$. It means that there exits a optimum value ${c^*}$ which satisfies: when $c$ is less than ${c^*}$, the utility of passengers is an increasing value of $c$; when $c$ is greater than ${c^*}$, the utility of passengers is an decreasing value of $c$.
  \item Passengers prefer greater departure frequency. The greater the $f$ is, the lager the passengers' utility will be.
\end{itemize}

In order to capture the concave property of utility function, we define its extreme point as $\frac{f}{{2\alpha }}$, which increases with the departure frequency. By the analysis above, we define
a utility function for passengers as

\begin{equation}
\begin{aligned}\label{equ:ECS6}
    {U_f} = cf - \alpha {c^2},
\end{aligned}
\end{equation}
in which $\alpha$ is a positive constant.
%
%

Then the mathematical description of optimization problem for passengers is as follows:

\begin{equation}
\begin{aligned}\label{equ:ECS7}
    &\mathop {\max }\limits_{f,c} {\rm{  }} \indent {U_f}\\
    &\rm{s}}{\rm{.t}}{\rm{.    }} \indent {f_{\min }} < f < {f_{\max }}{\rm{  }.
\end{aligned}
\end{equation}



%

\section{Stackelberg Game Approach}~\label{section: stackelberg equilibrium}
We first analyze the Stackelberg game process and obtain the closed form solutions to the game outcomes. On this basis, we will further prove the solution is the unique Stackelberg equilibrium.

\subsection{Stackelberg Game Process}
As discussed above, the operator determines the departure frequency, while the passengers choose the fare they are willing to pay and replan their travel routes. We model our problem as a Stackelberg game. The solution of the Stackelberg game is obtained using backward induction.

Given the players' feasible strategy and their utility function defined in~(\ref{equ:ECS4}) and~(\ref{equ:ECS7}), the game is played according to the following sequence.

\subsubsection{Operator first sets its strategy}First, the operator determines an initial departure frequency according to its cost constraints.
\subsubsection{Passengers choose the best response}
In the second stage, passengers know the decision of departure frequency made by operator. In order to improve their quality of experience (QoE), passengers make the best response to maximize their utility. Considering $U_f$ as function in $c$, we strive for the second order derivative $l\left( c \right)$ of~(\ref{equ:ECS4}) as

\begin{equation}
\begin{aligned}\label{equ:ECS9}
    l\left( c \right) = \frac{{{\partial ^2}{U_f}}}{{\partial {c^2}}} =  - 2\alpha<0.
\end{aligned}
\end{equation}

Since it is negative, it can be calculated directly that the utility of passengers is concave with a unique maximum. Then the first order derivative $h\left( c \right)$ can be computed as

\begin{equation}
\begin{aligned}\label{equ:ECS8}
    h\left( c \right) = \frac{{\partial {U_f}}}{{\partial c}} = f - 2\alpha c,
\end{aligned}
\end{equation}

Given the departure frequency $f$, and let $h\left( c \right) = 0$, we can get
\begin{equation}
\begin{aligned}\label{equ:ECS10}
    \frac{{\partial {U_f}}}{{\partial c}} = f - 2\alpha c = 0.
\end{aligned}
\end{equation}
%


Solving~(\ref{equ:ECS10}) for the unknown $c$, we have the best response of passengers

\begin{equation}
\begin{aligned}\label{equ:ECS11}
    {c^*} = c\left( f \right).
\end{aligned}
\end{equation}
\subsubsection{Operator resets its strategy based on the best response of passengers }
As a leader, the railway operator is aware of the fact that the passengers will choose their best response to its strategy. The operator tries to maximize its utility based on the best response of passengers.
Substituting the best response $c^*$ into~(\ref{equ:ECS4}), we can get the maximum operator's utility by
\begin{equation}
\begin{aligned}\label{equ:ECS12}
    \mathop {\max }\limits_{f,c} {\rm{ }}{U_l} = {\rm{ }}{{\rm{c}}^*}Q\left( {1 - {e_c}{c^*} + {e_f}f} \right) - Bf,
\end{aligned}
\end{equation}
in which
\begin{equation}
\begin{aligned}\label{equ:ECS13}
    B{\rm{ = }}\Omega T + 2{T_p}\left( {\phi  + RT + {S_0}\frac{T}{{{T_S}}}} \right).
\end{aligned}
\end{equation}

Then the problem boils down to solving the maximum value of (\ref{equ:ECS12}). We take the partial derivative of $f$, and let it be 0, as follows:
\begin{equation}
\begin{aligned}\label{equ:ECS14}
    \frac{{\partial {U_l}}}{{\partial f}} = \frac{\partial }{{\partial f}}\left[ {{{\rm{c}}^*}Q\left( {1 - {e_c}{c^*} + {e_f}f} \right) - Bf} \right] = 0.
\end{aligned}
\end{equation}

We can get the expression of ${f^*}$ from~(\ref{equ:ECS14}). $f^*$ is the most optimal solution to operator's departure frequency, which cannot only reduce the operating cost, but also provide higher-level quality of experience.

%
%

\subsection{Stackelberg Equilibrium}
In this section, the Stackelberg equilibrium of the game is defined. In addition, the solution ${c^*}$ of~(\ref{equ:ECS14}) and the solution ${f^*}$ of~(\ref{equ:ECS17}) is proved to be Stackelberg equilibrium of this game.

\emph{\textbf{Defination:}}~Let ${f^*}$ be a solution for~(\ref{equ:ECS4}) and ${c^*}$ is a solution for~(\ref{equ:ECS7}). For any $\left( {{f},{c}} \right)$, which satisfies ${f} \ge 0$ and ${c} \ge 0$, the point $\left( {{f^*},{c^*}} \right)$ is the Stackelberg equilibrium of this game if the following conditions are met:

\begin{equation}
\begin{aligned}\label{equ:ECS15}
{U_l}\left( {{f^*},{c^*}} \right) \ge {U_l}\left( {f,c} \right)\,
\end{aligned}
\end{equation}

\begin{equation}
\begin{aligned}\label{equ:ECS16}
{U_f}\left( {{f^*},{c^*}} \right) \ge {U_f}\left( {f,c} \right)\,
\end{aligned}
\end{equation}

\emph{\textbf{Theorem:}} There exits a unique Stackelberg equilibrium for the optimization mechanism proposed in this paper after a limited number of iterations.

\emph{\textbf{Proof:}} The Stackelberg equilibrium can be deduced by using backward induction. Firstly, we prove the existence of Stackelberg equilibrium for the passengers (follower). In the second stage of the game, with the announced strategy of the railway operator (leader), the passengers choose the best response strategy (the fare they can afford the service) according to~(\ref{equ:ECS11}) so as to maximize $U_f$. According to~(\ref{equ:ECS6}), we can know that $U_f$ is a strictly concave function of $c$, as the second order derivative $U_f^{''} =  - 2\alpha  < 0$ where $\alpha  > 0$. Thus, the decision made by passengers has a unique solution, with an offered departure frequency $f$.
\begin{figure*}[!t]
\begin{minipage}[t]{0.33\textwidth}
  \centering
  \includegraphics[width=2.5in]{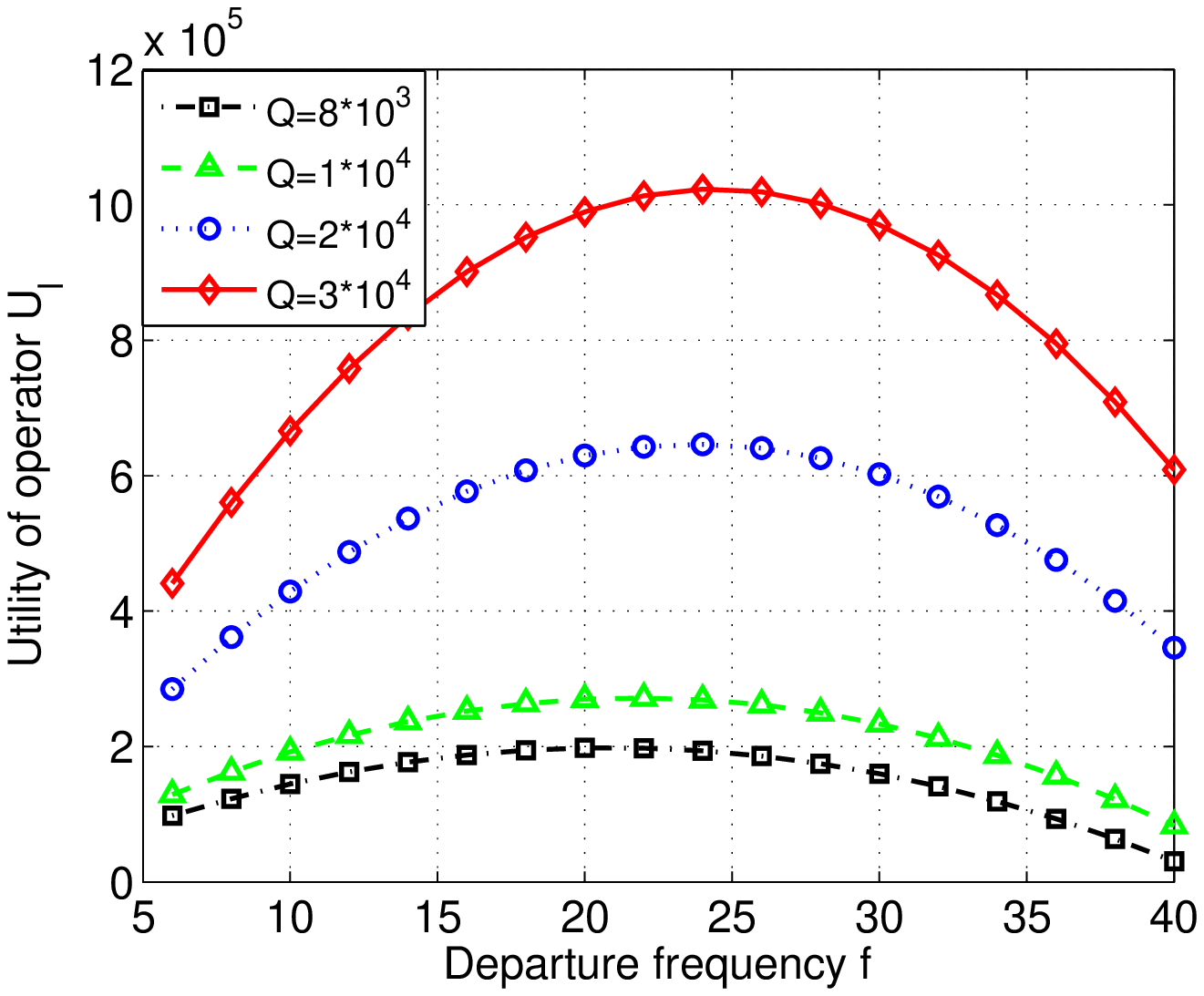}
  \caption{Utility of operator vs.\protect\\ departure frequency}
  \label{frequencyUl}
\end{minipage}
\begin{minipage}[t]{0.33\textwidth}
  \centering
  \includegraphics[width=2.5in]{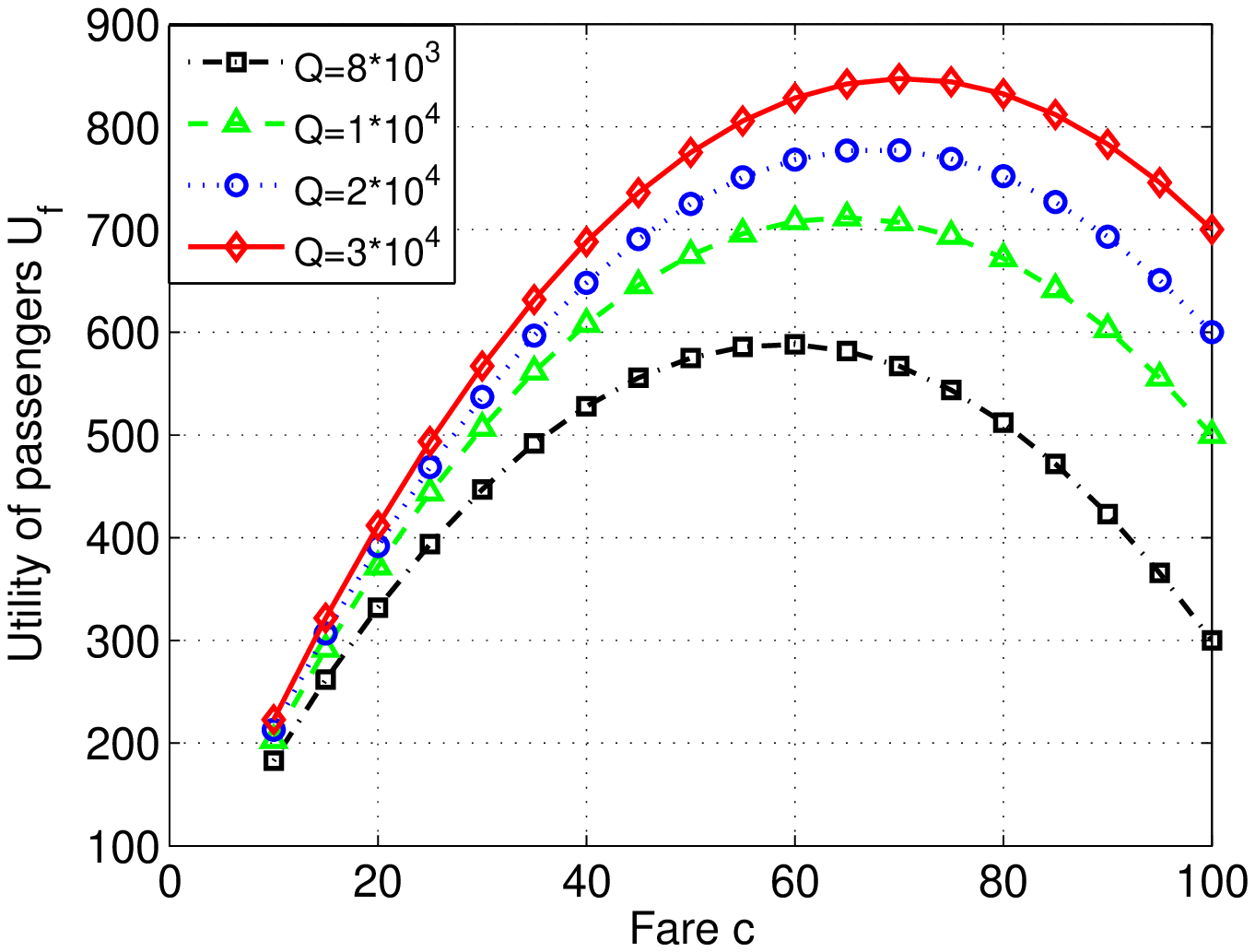}
  \caption{Utility of passengers vs.\protect\\ fare (given departure frequency)}
  \label{optimalfare}
\end{minipage}
\begin{minipage}[t]{0.33\textwidth}
  \centering
  \includegraphics[width=2.5in]{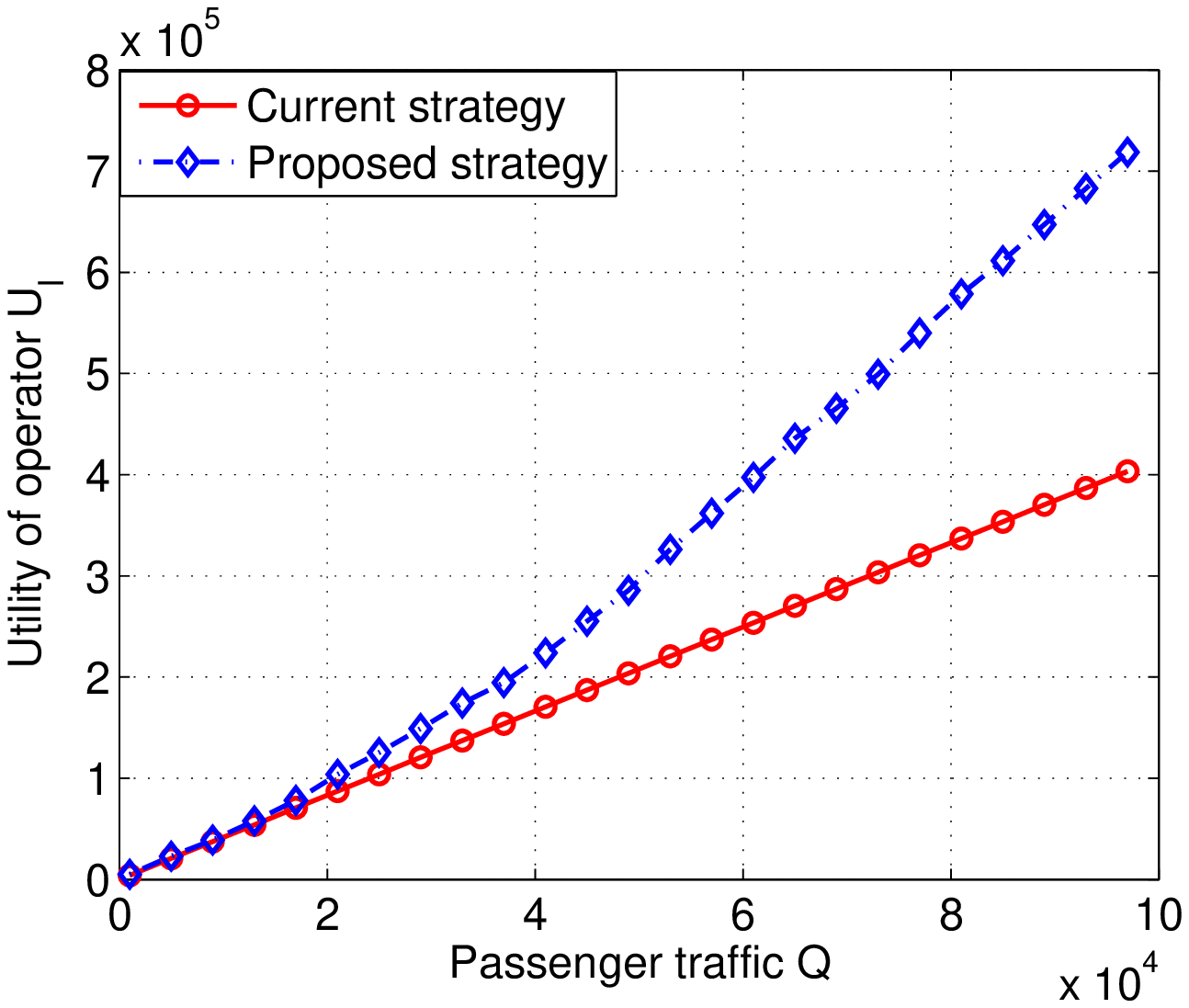}
  \caption{Comparison between the proposed scheme and baseline strategy}
  \label{compare}
\end{minipage}
\end{figure*}

Secondly, we prove the the existence of Stackelberg equilibrium for the railway operator (leader). Since the operator is aware of the decision made by passengers in the second stage according to Stackelberg game theroy, we first discuss the decision made by the passengers. As discussed above, with a given frequency $f$, passengers determine the fare $c$
they wish based on the following equation
\begin{equation}
\begin{aligned}\label{equ:ECS17}
    \frac{{\partial {U_f}}}{{\partial c}} = 0,
\end{aligned}
\end{equation}
the best response strategy made by passengers is described as
\begin{equation}
\begin{aligned}\label{equ:ECS18}
    {c^*} = min\left\{ {\frac{f}{{2\alpha }},{c_{\max }}} \right\}.
\end{aligned}
\end{equation}

With the information of $c^*$, we can rewrite the utility of operator $U_l$ as
\begin{equation}
\begin{aligned}\label{equ:ECS19}
    {U_l} = {c^*}\left( {1 - {e_c}{c^*} + {e_f}f} \right) - Bf.
\end{aligned}
\end{equation}

According to~(\ref{equ:ECS3}), we get the second derivation $U_l^{''} \le 0$. Hence, $U_l$ is is strictly convex with respect to $f$. Thus, there exists a unique Stackelberg equilibrium.

%

\section{Numerical Experiments}\label{section: numerical experiments}
This section evaluates the performance of the proposed Stackelberg-based optimization method with MATLAB. The tool of data processing we used is statistical product and service solutions (SPSS), which is widely used for statistical analysis, data management and data document.

\subsection{Data Statistics and Analysis}
The numerical simulations are based on the data from Nanjing Metro. We extract the passenger traffic data of one rail line at different time of the day (every hour is divided as one time terminal throughout the running time of Nanjing transit subway), and we adopt the data of ten days (data of the second quarter of 2010), part of which is shown in Table~\ref{traffic}. Based on the traffic data of all the stations, we can get the passengers arrival rate curve, which is one of the input parameters of the game model.

\begin{table}[!ht]
\renewcommand{\arraystretch}{1.2}
\caption{Passenger Traffic Examples}
\label{traffic}
\centering
\begin{tabular}{cccccc}
\hline
Times Interval& ${n_{in}}$ & ${n_{out}}$  & Times Interval& ${n_{in}}$ & ${n_{out}}$\\
\hline
05:00-06:00 & 1138  & 47     & 14:00-15:00& 26235 & 26875 \\
06:00-07:00 & 7893  & 4316   & 15:00-16:00& 26672 & 25052 \\
07:00-08:00 & 19080 & 15105  & 16:00-17:00& 28379 & 25499 \\
08:00-09:00 & 25993 & 28353  & 17:00-18:00& 32113 & 28961 \\
09:00-10:00 & 24706 & 27040  & 18:00-19:00& 25217 & 27842 \\
10:00-11:00 & 23736 & 26265  & 19:00-19:00& 17624 & 18203 \\
11:00-12:00 & 23045 & 24807  & 20:00-21:00& 18471 & 15594 \\
12:00-13:00 & 23168 & 23429  & 21:00-22:00& 16579 & 14945 \\
13:00-14:00 & 25820 & 25554  & 22:00-23:00& 10543 & 11228 \\
\hline
\end{tabular}
\end{table}


The total passengers traffic can be calculated as
\begin{equation}
\begin{aligned}\label{equ:ECS20}
    Q = \sum\limits_{i = 1}^m {\int\limits_{{t_1}}^{{t_2}} {{\lambda _i}\left( t \right)dt} },
\end{aligned}
\end{equation}
where ${\lambda _i}\left( t \right)$ is passenger arrival rate at the station $i$.
\subsection{Simulation Set-up and Performance Analysis}
According to the train characteristics of the Nanjing Metro Line 1, the capacity of each train is 1860 passengers (6 persons per square meter), and the minimum headway $h_{min}$ between two successive trains is 90 s. The settings of relevant parameters can be found in Table~\ref{parameters}.

\begin{table}[!htp]
\renewcommand{\arraystretch}{1.2}
\caption{Simulation Parameters}
\label{parameters}
\centering
\begin{tabular}{cccc}
\hline
Parameters &  Value & Parameters & Value\\
\hline
$h_{min}$  &  90   & $h_{max}$  &  600    \\
$e_c$      &  0.01 & $e_f$      &  0.01   \\
$\Omega$   &  3000 & $R$        &  1500   \\
$T_p$      &  1    &$\theta$    &  0.5 $\sim$ 0.8  \\
\hline
\end{tabular}
\end{table}

By applying the proposed method, we obtain the optimal departure frequency for each time period.
The complicated interactions between operator and passengers are analysed in Fig.~\ref{frequencyUl} and Fig.~\ref{optimalfare}. In the first stage of the game, the operator makes the optimal decision according to the passenger traffic. Given departure frequency, the passengers choose their optimal response in the second stage of the game. For example, as is shown in Fig.~\ref{frequencyUl}, the optimal frequency is 24 times per hour when the passenger traffic is $3\times10^4$. Given the frequency to be 24 times per hour, the optimal response of passengers is 70, which is shown in Fig.~\ref{optimalfare}. We choose 0.05 as the conversion coefficient. Correspondingly, the fare that passengers would like to pay is 3.5.


%

Fig.~\ref{compare} shows the comparison between the current strategy and the proposed optimal strategy. The proposed method can improve the utility of operator by over 40.67\% during rush hours (passenger traffic is about $3\times10^4$) and 8.33\% during non-rush hours (passenger traffic is about $1\times10^4$).


%
%

From the simulation results, we can see that our proposed scheme has obvious advantages. This will not only meet passengers' demand, but also maximize operate income as well as passengers' quality of experience, and eventually improve the systematic efficiency.

\section{Conclusion} \label{section: conclusion}
The main contribution of this paper is applying Stackelberg game theory to the dynamic decision-making process of departure frequency. The aim is to alleviate the perceived conflict between passengers' and operator's objectives. In detail, the game is played as: the railway operator first sets its strategy, i.e., an initial departure frequency, then the passengers choose the best response strategy to maximize their utility. After knowing the best response of passengers, operator resets its strategy. In the model establishment and solving process, we give the utility functions of both sides as the game model input and obtain the closed form solutions to the game model output. In addition, the proposed method is proved to be feasible and efficient by the simulation works. The optimized departure frequency can effectively improve the resource utilization, the traffic efficiency and the QoE of passengers during rush hours and non-rush hours.

%
%

\clearpage
\end{document}